# Assessing the Human-Likeness of LLM-Driven Digital Twins in Simulating Health Care System Trust

Yuzhou Wu[1,2], Mingyang Wu[1,3], Di Liu[1,3,4], Rong Yin[1,3,4*], Kang Li[1,4*]

[1]West China Biomedical Big Data Center, West China Hospital, Chengdu, 610041, China

[2]School of Industrial and Operations Engineering, University of Michigan, Ann Arbor, 48109, US

[3]Department of Industrial Engineering, Sichuan University, Chengdu, 610207, China

[4]Med-X Center for Informatics, Sichuan University, Chengdu, 610041, China

[*]These authors contributed equally to this work.

## Abstract

Serving as an emerging and powerful tool, Large Language Model (LLM)-driven Human Digital Twins are showing great potential in healthcare system research. However, its actual simulation ability for complex human psychological traits, such as distrust in the healthcare system, remains unclear. This research gap particularly impacts health professionals' trust and usage of LLM-based Artificial Intelligence (AI) systems in assisting their routine work. In this study, based on the Twin-2K-500 dataset, we systematically evaluated the simulation results of the LLM-driven human digital twin using the Health Care System Distrust Scale (HCSDS) with an established human-subject sample, analyzing item-level distributions, summary statistics, and demographic subgroup patterns. Results showed that the simulated responses by the digital twin were significantly more centralized with lower variance and had fewer selections of extreme options (all p<0.001). While the digital twin broadly reproduces human results in major demographic patterns, such as age and gender, it exhibits relatively low sensitivity in capturing minor differences in education levels. The LLM-based digital twin simulation has the potential to simulate population trends, but it also presents challenges in making detailed, specific distinctions in subgroups of human beings. This study suggests that the current LLM-driven Digital Twins have limitations in modeling complex human attitudes, which require careful calibration and validation before applying them in inferential analyses or policy simulations in health systems engineering. Future studies are necessary to examine the emotional reasoning mechanism of LLMs before their use, particularly for studies that involve simulations sensitive to social topics, such as human-automation trust.

**Keywords:** Large Language Models, digital twins, health care system, distrust

## 1. Introduction

Distrust in the health care system is widely considered an important psychological factor of patient behavior, including service utilization, adherence, information disclosure, and health outcomes, particularly among minority, low-income, and historically marginalized populations [1], [2]. The Health Care System Distrust Scale (HCSDS), developed by Rose et al. [3], provided a structural and well-validated measuring tool, as well as showing great reliability and validity among diverse groups. In the world of health systems engineering, the ability to measure and simulate distrust is a practical necessity. Getting this right can help hospitals fine-tune their services, guide the design of public health campaigns, and support the development of fairer health policies [4], [5].

Recently, Artificial Intelligence (AI), particularly Large Language Models (LLMs), has shown great potential in the healthcare sector. They can help answer patients' questions, simulate behaviors, serve as virtual subjects, and carry out predictive analytics [6], [7], [8]. However, when LLMs are asked to, for example, fill out surveys or mimic human responses, they are more likely to choose mid-scale options. Meanwhile, the overall spread of answers is narrower than what we see in real human beings [9], [10], [11]. Moreover, LLMs struggle to capture demographic heterogeneity, such as differences in race or education [12], [13]. The objective of this study is to construct and test digital twins and LLMs in reflecting real human behaviors and responses. We aim to quantitatively assess item-level concordance between LLM-generated and real-world data, investigating the LLM-driven human digital twin's ability to reproduce demographic heterogeneity, and giving suggestions for calibration and validation of LLM-based digital twins in health systems engineering.

## 2. Methods

To build the human digital twins, we used the Twin-2K-500 dataset [14]. It contains detailed profiles of 2058 U.S. adults as personas. For each persona, there are more than 500 features, covering everything from basic demographics to psychological, economic, and behavioral measurement results. Then we use ChatGPT-4 to power the digital twins, where the persona summaries provided in the dataset were used as prompts, ensuring that it closely reflects the human diversity within the dataset. The human reference sample is drawn from 400 Philadelphia jurors in the study of Rose et al. [3].

To ensure comparability with the human subject study, we applied stratified random sampling to obtain a 500-case subsample from the full dataset (N = 2,058). Stratification variables included gender, age group, and ethnicity. Moreover, education levels were treated as a secondary variable due to the limitation in the full dataset. Due to the differences between the dataset and the human subject study, age groups were aggregated into three groups: 18-30, 31-50, and 51+. Target sample sizes for each stratum were proportionally scaled to 500 cases. If a stratum was underrepresented, all available cases were included, and remaining slots were randomly filled. The final subsample matched the reference human population in major demographics to a high degree. However, due to data source limitations, the educational distribution showed some deviation (See Table 1), with around 70% of the samples having received undergraduate or higher education.

**Table 1.** Demographic Characteristics Comparison

| Variable | Category | Reference (N=400) n (%) | Subsample (N=500) n (%) |
|---|---|---|---|
| **Gender** | Female | 244 (61.0%) | 304 (60.8%) |
| | Male | 147 (36.8%) | 196 (39.2%) |
| **Age group** | 18–30 | 105 (26.3%) | 130 (26.0%) |
| | 31–50 | 189 (47.3%) | 222 (44.4%) |
| | 51+ | 92 (23.0%) | 148 (29.6%) |
| **Race/Ethnicity** | African American | 156 (39.0%) | 192 (38.4%) |
| | White | 178 (44.5%) | 240 (48.0%) |
| | Hispanic | 16 (4.0%) | 25 (5.0%) |
| | Asian | 8 (2.0%) | 11 (2.2%) |
| | Other | 24 (6.0%) | 32 (6.4%) |
| **Education** | ⩽ High school | 111 (27.8%) | 85 (17.0%) |
| | ⩾ Some college | 280 (70.0%) | 415 (83.0%) |

***Note:*** *There is a small proportion of missing values (1.4% to 2.8%) in the human reference data*

In this study, we applied the 10-item Health Care System Distrust Scale (HCSDS) developed by Rose et al. [3] with 5-point Likert responses. Three positively worded items, B ("My medical records are kept private"), H ("I receive high-quality medical care from the health care system"), and I ("The health care system puts my medical needs above all other considerations"), were reverse-scored to measure distrust. For each item, we calculated mean values, standard deviations, and drew distribution histograms for visualization. Chi-square tests were conducted to compare response distributions between digital twins and human reference. Moreover, we conducted between-group differences analyses and subgroup analyses. All analyses were conducted in R 4.4.1 with a significance level of $\alpha = 0.05$.

## 3. Results

The digital twins tended to give answers that clustered tightly around the center, resulting in taller peaks on the response curves. By comparison, real participants showed a broader range of scores, with more responses spread out and the peaks noticeably lower (See Figure 1).

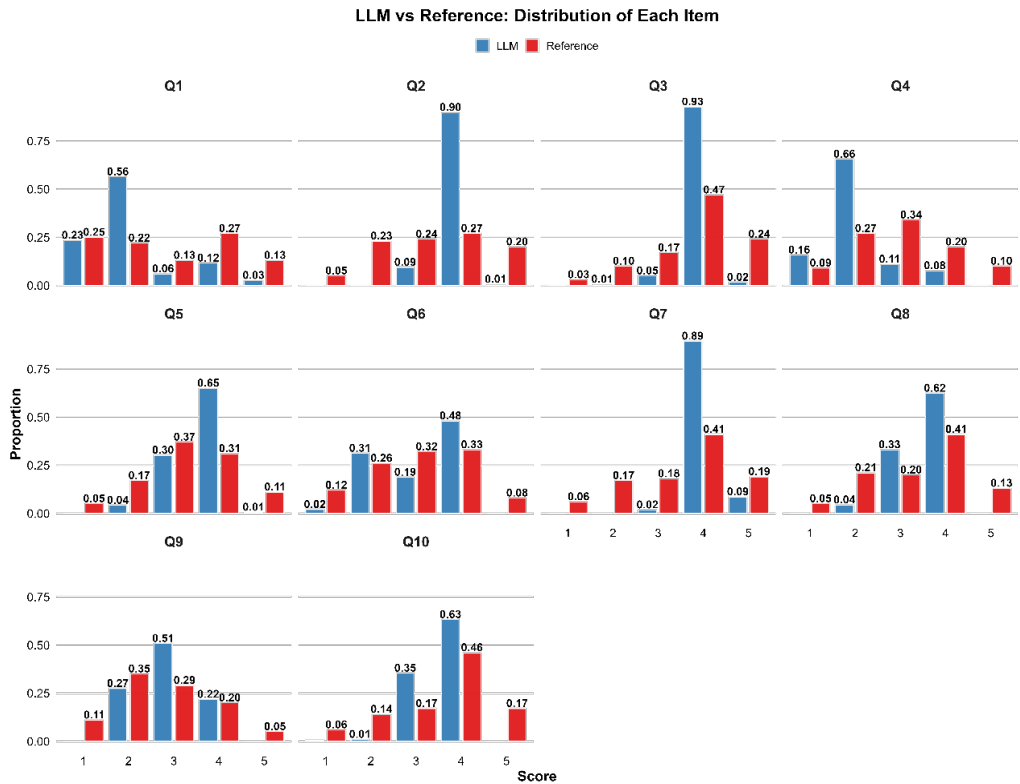

**Figure 1.** Item-wise Response Distributions of LLM vs Reference

To quantitatively evaluate the correspondence between the digital twin results and the human reference, we conducted chi-square tests for each item. The results showed the highly significant differences appearing in all items (all p<0.001), with chi-square statistics ranging from 158.43 to 996.73. It indicates that LLM-driven digital twins exhibit systematic response bias compared to real humans.

Table 2 lists the item-wise results in both the LLM simulation and the human reference survey, with items B, H, and I reverse-scored to measure distrust. The responses from the digital twins are relatively centralized with a smaller standard deviation compared with the human reference ones.

For the effect size, Cohen's d values of items H and I (d = -1.44 and -1.38) are considerably large, indicating that digital twins generate significantly lower distrust compared to humans on these items. Moderate to large effect sizes are also observed for items B, D, and G, further confirming the differences between the digital twins and actual human responses.

**Table 2.** Item-level Statistics Comparison

| Item | LLM Mean | LLM SD | Reference Mean | Reference SD | Cohen's d |
|------|----------|--------|----------------|--------------|-----------|
| A | 2.14 | 0.99 | 2.81 | 1.40 | -0.57 |
| B | 1.90 | 0.32 | 2.63 | 1.18 | -0.89 |
| C | 3.95 | 0.30 | 3.82 | 1.02 | 0.19 |
| D | 2.10 | 0.75 | 2.95 | 1.11 | -0.91 |
| E | 3.62 | 0.58 | 3.29 | 1.03 | 0.41 |
| F | 3.13 | 0.93 | 3.32 | 1.23 | -0.18 |
| G | 4.06 | 0.33 | 3.53 | 1.15 | 0.66 |
| H | 1.42 | 0.58 | 2.64 | 1.10 | -1.44 |
| I | 2.06 | 0.70 | 3.27 | 1.06 | -1.38 |
| J | 3.61 | 0.53 | 3.54 | 1.11 | 0.09 |

A subgroup analysis was conducted to assess the human digital twins' ability to model demographic heterogeneity in terms of age, gender, and educational levels, comparing them with the human reference data. In both the LLM and the human sample, no significant differences were observed in these three subgroups; however, higher education levels are associated with slightly higher distrust scores. As a result, the digital twins reproduced the main demographic homogeneity seen in human reference.

## 4. Discussion

This study is the first to systematically compare the performance of the LLM-driven human digital twins with the authoritative human sample data by Rose et al. on the Health Care System Distrust Scale (HCSDS). We found that the digital twins' responses are significantly more concentrated, with a lower standard deviation, compared to the human reference. Moreover, the rates of choosing extreme options are relatively low. These findings are highly consistent with international research on "variance collapse" and "extreme aversion" in AI behavioral simulation [15], [16]. As for the mechanisms behind such behaviors, firstly, most training data of LLMs may be mainstream and moderate narratives, which means LLMs are simply less exposed to those extreme views, so that they may be difficult to mimic extreme responses. Secondly, the safety mechanisms of LLMs themselves may filter out controversial or sensitive content, which further reduces extreme outputs. [17].

The subgroup analysis suggests that LLM-driven digital twins can generally reproduce demographic homogeneity in gender and age. Nevertheless, it is not sensitive enough to the subtle differences related to areas such as education, echoing the weak positive relationship between educational levels and distrust reported in Rose et al.'s study. The strength of the Twin-2K-500 dataset lies in its broad coverage of persona features, providing us with a solid foundation for the construction and testing of this personalized digital-twin simulation. LLMs can be used for survey piloting or serving as virtual participants [18], [19]. However, LLMs struggle to mimic minority group attitudes and extreme responses on sensitive social topics such as health care system distrust [6], [20].

Future studies can try to use strict calibration, postprocessing such as importance reweighting or parameter tuning, and manual review to improve the reliability and representativeness of LLM-driven human digital twins in the healthcare sector. In addition, integrating item response theory (IRT), multi-model benchmark tests, and cross-cultural data can help clarify the limitations of LLM-driven human digital twins under various healthcare scenarios.

## 5. Conclusion

This study finds that LLM-driven human digital twins exhibit variance collapse and a marked aversion to extreme responses on the Health Care System Distrust Scale (HCSDS) compared with human reference. The score distribution of all items is significantly different from that of real human participants. Although LLMs can reproduce major demographic patterns in age and gender, their ability to capture subtle, heterogeneous differences remains limited. Therefore, we suggest it is essential to carry out calibration and manual review before using LLM-powered score distribution simulation.

## Acknowledgement

Part of this study was supported by the Natural Science Foundation of Sichuan, China (2025NSFSC1985); Part of this study was supported by the 1·3·5 project for disciplines of excellence, West China Hospital, Sichuan University (ZYYC21004).